\documentclass{jfm}

\usepackage[latin1]{inputenc}
\usepackage[english]{babel}
\usepackage{float}
\usepackage{subfig}
\usepackage{textcomp}
\usepackage{amsmath}
\usepackage{graphicx}
\usepackage{setspace}
\usepackage{esint}
\usepackage{natbib}
\usepackage{color,soul}

\usepackage{psfrag}

\usepackage[T1]{fontenc}               
\usepackage[section]{placeins}         

\usepackage{fancyhdr}
\usepackage{mathrsfs}
\usepackage{amsfonts}
\usepackage{lscape}
\usepackage{longtable}

\usepackage{array}
\newcolumntype{L}[1]{>{\raggedright\let\newline\\\arraybackslash\hspace{0pt}}m{#1}}
\newcolumntype{C}[1]{>{\centering\let\newline\\\arraybackslash\hspace{0pt}}m{#1}}
\newcolumntype{R}[1]{>{\raggedleft\let\newline\\\arraybackslash\hspace{0pt}}m{#1}}

\ifCUPmtlplainloaded \else
  \checkfont{eurm10}
  \iffontfound
    \IfFileExists{upmath.sty}
      {\typeout{^^JFound AMS Euler Roman fonts on the system,
                   using the 'upmath' package.^^J}%
       \usepackage{upmath}}
      {\typeout{^^JFound AMS Euler Roman fonts on the system, but you
                   dont seem to have the}%
       \typeout{'upmath' package installed. JFM.cls can take advantage
                 of these fonts,^^Jif you use 'upmath' package.^^J}%
      }
  \else
  \fi
\fi

\ifCUPmtlplainloaded \else
  \checkfont{msam10}
  \iffontfound
    \IfFileExists{amssymb.sty}
      {\typeout{^^JFound AMS Symbol fonts on the system, using the
                'amssymb' package.^^J}%
       \usepackage{amssymb}%
         \let\leq=\leqslant
         
      }{}
  \fi
\fi

\ifCUPmtlplainloaded \else
  \IfFileExists{amsbsy.sty}
    {\typeout{^^JFound the 'amsbsy' package on the system, using it.^^J}%
     \usepackage{amsbsy}}
    {\providecommand\boldsymbol[1]{\mbox{\boldmath $##1$}}}
\fi

\newsavebox{\astrutbox}
\sbox{\astrutbox}{\rule[-5pt]{0pt}{20pt}}

\title[Stability of the laminar boundary layer beneath a Stokes wave]{On the stability of the laminar boundary layer beneath a Stokes wave}

\author[F. Fedele] 
{Francesco Fedele$^1$ \thanks{Email address for correspondence: fedele@gatech.edu}}   

\affiliation{$^1$School of Civil and Environmental Engineering, Georgia Institute of Technology, Atlanta, GA 30332, USA}

\pubyear{2018}
\volume{xxx}
\pagerange{xxx--xxx}
\date{?; revised ?; accepted ?. - To be entered by editorial office}

\begin{document}
\maketitle 
\begin{abstract}
The linear stability of the laminar boundary layer flow of a Stokes wave in deep waters  is investigated by means of a 'momentary' criterion of instability for unsteady flows~\citep{Blondeaux1979}. In the parameter range investigated, it is found that the flow is stable to 2-D perturbations. The least stable eigenmode of the resulting Orr-Sommerfield spectrum attains its maximum beneath the boundary layer of the Stokes wave.  Moreover, an analysis of the associated pseudospectrum indicates that the laminar flow when modified by imperfections is unstable due to the non-normality of the Orr-Sommerfeld operator, and the unstable pseudo-eigenmodes tend to peak within the boundary layer. The laminar flow of the Stokes wave is also stable to 3-D streamwise-independent perturbations. Instability is observed for the laminar flow with imperfections. The associated unstable pseudo-eigenmodes are streamwise vortical rolls similar to Langmuir cells. The laminar boundary layer flow of a Stokes wave appears to be stable to infinitesimal perturbations, but it may likely be unstable to finite perturbations, as in Poiseuille pipe flows.  The present results are supportive of the recent experimental evidence of spontaneous occurrence of turbulence beneath unforced non-breaking surface waves. 
\end{abstract}


\section{Introduction}
Surface waves generate mixing in the upper ocean through breaking and Langmuir turbulence~\citep{Melville1996,mcwilliams1997,belcher2012}. Recent experimental and numerical studies also suggest another source as the spontaneous occurrence of turbulence beneath surface waves in absence of breaking and wind forcing~\citep{Babanin2009,Babanin2010,Babanin2010a,Chalikov2012,Chalikov2012a, Onorato2019}. In particular, these studies foreshadow that the oscillatory viscous flow beneath unforced non-breaking waves becomes unstable and generates vorticity injected into the subsurface layers. In this regard, the theoretical work by~\cite{Benilov2012} proposes an inviscid vortex instability of the potential surface waves, where the interaction between the vortex and potential motions cause the vertical transport of the momentum injecting vorticity beneath waves. 
\cite{Babanin2009} conjectured a viscous instability of the orbital motion beneath a non-breaking wave that can become unstable, with the laminar flow beneath the wave surface transitioning to turbulence at a critical wave Reynolds number $Re_w\simeq3000$. Here, $Re_w=U a/\nu=a^2\sigma/\nu$ where $U=a\sigma$ is the characteristic orbital velocity, $a$ and $\sigma$ are wave amplitude and frequency, and $\nu$ the kinematic viscosity coefficient. 
Moreover, numerical simulations by~\cite{Chalikov2012} indicate that vorticity and turbulence generate in the vicinity of non-breaking wave crests and then spread downward. 

Wave-tank measurements of the velocity field of oscillatory flows induced by mechanically generated random waves indicate that at $Re_w\simeq10^4$ the associated spectrum and structure functions are consistent with a Kolmogorov inertial cascade~\citep{Onorato2019}. However, when \cite{Banner2012} reproduced the experiments of~\cite{Babanin2009} using conventional dye techniques they did not observe any turbulent mixing over the short fetches near the wave generator, so to capture a Stokes wave before modulational instability initiates. In particular, the dye did not disperse in the wave boundary layer up to $Re_w\simeq7000$, where breaking appeared. \cite{Chalikov2012} argued that for turbulent mixing to occur requires long time scales of $O(10^3)$ wave periods in non-steep short waves. On that basis, mixing would not be expected in~\cite{Banner2012}'s experiments, where a variety of waves were generated with gentle to steep slopes up to breaking.

The abovementioned experimental studies on wave-induced turbulence motivate this investigation of the linear stability of the laminar boundary layer beneath a Stokes wave in deep waters, a free-shear stress surface.  The associated viscous flow is unsteady as it inherits the oscillatory nature of the Stokes wave. Moreover, the boundary layer is weaker than that at the bottom of a Stokes wave~(\cite{Phillips1977}, section 3.4). This may be the reason of why further studies of its stability have not been reported. In contrast, the stability of  oscillating boundary layers at the flat bottom of progressive or solitary waves has been the subject of several studies~\citep{blondeaux_pralits_vittori2012,vittori_blondeaux2008,Blondeaux1979,Hall1978}. In particular, \cite{Blondeaux1979} found that these flows are linearly unstable based on a 'momentary' criterion for instability, as an alternative to Floquet theory~(see, for example, \cite{Henningson2001,Fedele_pipe2005}).  In particular, the authors assume that the infinitesimal disturbance evolves on faster time scales and shorter space scales than those of the unsteady base flow. As a result, for large Reynolds numbers the fast dynamics of the disturbance uncouples from the slower evolution of the unsteady base flow, and the 'momentary' instability of the base flow can be probed at a given point in space or at an instant of time. This is reflected  in the slow time and space variables both appearing as parameters in the Orr-Sommerfeld (O-S) equation governing the fast evolution of the disturbance. 

Moreover, the non-normality of the O-S operator suggests that the viscous boundary layer of a Stokes wave~(\cite{Phillips1977}, section 3.4) can be nonlinearly unstable to finite perturbations, similarly to the wall boundary layer of Couette or Poiseuille pipe flows~\citep{Trefethen1999,Henningson2001}. Indeed, linear stability theory suggests that pipe flows remain laminar for all Reynolds numbers $Re$, but the first indication of transition to turbulence in experiments  occurs near $Re\approx1600$~\citep{Willis2008,Eckhardt2018} and the flow is fully turbulent beyond $Re\approx2900$~\citep{Barkley2005}. Following the seminal work of \cite{Waleffe1995,WaleffePOF1997} on 3-D~Self-Sustaining Processes (SSPs), \cite{FaisstEckhardt} and \cite{HofScience2004} discovered the existence of finite amplitude traveling waves (TW) in pipe flows (see also~\cite{FedeleFDR2012,FedeleDutykhEPL2013}), which are exact solutions of the Navier-Stokes equations, suggesting new routes for transition to turbulence. 
SSPs are generic features of shear flows, such as Couette or Poiseuille pipe flows. They are linearly stable at any Reynolds numbers. To induce instability, finite-amplitude streamwise rolls like the least stable O-S eigenmode are added to the base laminar flow. At this initial stage, the rolls remodulate the streamwise velocity spanwise in the form of "streaks" by way of a  redistribution of the momentum in the crossflow planes. The streaks (base flow + rolls) are linearly unstable to spanwise disturbances. The instability leads to a streamwise ondulation, which regenerates the streamwise rolls via a nonlinear self-interaction, and the process repeats sustaining itself. 



There is a similarity between turbulent wall boundary layers and the upper layers within wind-driven ocean surface waves. Notebly, both display surprisingly similar coherent structures in the form of streamwise streaks align with the wind.  In the ocean, these streaks are attributed to the subsurface vortex motions induced by Langmuir cells or rolls~\citep{craik_leibovich1976,leibovich_1977}. In particular, a wind stress applied to an initially flat ocean surface generates a drift current in a thin layer near the surface. Part of the drift  arises from the second order Stokes streaming purely due to the created wave motion. However, the major contribution of the drift is due to direct momentum transfer from the wind to the waves, at least for short fetches~\citep{craik_leibovich1976,leibovich_1977,leibovich_radhakrishnan_1977,leibovich1980}. 

Interestingly, \cite{leibovich_1977} in Figure 1 of his paper provides a qualitative sketch and description of the Langmuir circulation that anticipates \cite{Waleffe1995}'s SSP structures: "\textit{The wave pattern advances in the wind direction, creating a drift that is second order in wave slope. The drift is larger where the excursion of the water from the mean surface is greater, creating the possibility of an undulating wave drift. Vortex lines in the frictional current are distorted producing a streamwise vorticity component and mixing due to the induced vertical motions.This mixing causes a feedback, and alters the frictional current.}"


In the present work, we shall investigate the linear stability of the unsteady laminar boundary layer beneath a Stokes wave applying a 'momentary' criterion of instability~\citep{Blondeaux1979}. The present study will contribute to shed more light into the similarity of Langmuir structures and Waleffe's SSPs and their role into the transition to turbulence in unforced non-breaking waves. 
In our search for unstable modes, we first focus on the stability of two-dimensional (2-D) perturbations because the Squire theorem guarantees that a shear flow becomes unstable to 2-D wavelike perturbations at a Reynolds number that is smaller than any value for which unstable 3-D disturbances exist~(see, e.g.~\cite{Henningson2001}). We then study the stability of the Stokes flow to 3-D streamwise-independent perturbations. 


The paper is structured as follows. In section 1 we revisit the generation of vorticity at a free surface. Then, in section 2 we derive the Orr-Sommerfeld equation for 2-D perturbations and the associated spectrum is studied in section 3. The sensitivity of the eigenvalues to finite defects or imperfections added to the base flow is investigated by means of the pseudospectrum in section 4. In section 5 the stability to 3-D streamwise independent perturbations is presented.  Finally, we conclude by discussing the relevance of our results to wave-induced turbulence by unforced non-breaking waves.   

\section{Vorticity generated at a zero-stress free surface}

In this section we briefly review the mathematical description of vorticity generated on a one-dimensional (1-D) zero-stress free surface.  In general, vorticity is generated at free surfaces whenever there is flow within regions of surface curvature~\citep{Wu1995,Lundgren}. A theory for steady surfaces was formulated by~\cite{Longuet_Higgins_curvature1998} and extended to unsteady surfaces by~\cite{Fedele2016}.
The condition of zero interfacial shear stress determines the strength of the
vorticity at the surface. Consider the two-dimensional  (2-D) frame $(x,z)$, where $x$ is the horizontal direction and $z$ is the vertical. For the special case of unidirectional waves propagating along $x$ and surface displacements $z=\eta(x,t)$, the vorticity vector created on the surface is orthogonal to the $xz$~plane  and it is given by~\citep{Fedele2016}
\begin{equation}
\omega_s=-\frac{2}{1+\eta_{x}^2}\left(\eta_{xt}+u\eta_{xx}\right),
\label{Omega}
\end{equation}
where $u$ is the horizontal particle velocity at the surface and the subscript in $\eta_x$ denotes derivative, i.e. $\partial_x\eta$. For a steady wave $\eta(X)=\eta(x-ct)$ traveling at the phase speed $c$, the above expression simplifies to~\citep{Longuet_Higgins_curvature1998,Longuet_Higgins_JFM_bores}
\begin{equation}
\omega_{s}=-2K(u-c),
\label{Omega1}
\end{equation}
where $u-c$ is the orbital velocity in the comoving frame $X$ traveling at speed $c$, and  
\[
K=\frac{\eta_{XX}}{\left(1+\eta_{X}^2\right)^{3/2}}
\]
is the surface curvature. At crests, where $u<c$ and $K<0$, vorticity is negative or clockwise and it vanishes when $u=c$~\citep{Fedele2016}. 

This non-zero vorticity resides in a vortex sheet along the
free-surface even for the ideal flow field beneath irrotational waves~\citep{Longuet_Higgins_curvature1998}. However, in realistic flows molecular viscosity is effective in diffusing the surface vorticity deep into the fluid and a laminar boundary layer is formed by the balance of advection and viscous effects. The diffusion-advection process obeys the 2-D vorticity equation 
\begin{equation}
\partial_{t}\omega+\mathbf{U}\bullet\nabla\omega=\nu\nabla^2\omega\quad z\leq\eta,\label{vor}
\end{equation}
with boundary conditions $\omega=\omega_s$ at the surface $z=\eta$~(see Eq.~\eqref{Omega}) and vanishing vorticity at infinite depths,  and $\nabla=(\partial_x,\partial_z)$. 

In the following, we will show that the Eulerian velocity field $\mathbf{U}$ consists of an irrotational component due to the Stokes wave and a rotational part induced by viscous effects. Clearly, vortex stretching and tilting are absent since the flow is planar.


\section{2-D Stability of the laminar Stokes boundary layer}
Consider a linear Stokes wave of amplitude $a$ traveling in deep waters along $x$ at the speed $c=\sigma/k$, where  $k$ is the wavenumber and  $\sigma=\sqrt{gk}$ is the associated frequency and $g$ is the gravitational acceleration. Correct to $O(a)$, the surface displacements $\eta$ and the components of the induced irrotational flow field $\mathbf{U}_0=(U_0,W_0)$ are given by:
\begin{equation}
\eta(x,t)=a\cos\theta,
\end{equation}  
where $\theta=kx-\sigma t$, and
\begin{equation}
U_0(x,z,t)=U\exp(kz)\cos\theta,\quad W_0(x,z,t)=U\exp(kz)\sin\theta,\label{Ui}
\end{equation}  
where $U=a\sigma$ is the characteristic wave-induced orbital velocity.

Consider the following perturbation expansion for the Eulerian velocity field $\mathbf{U}$ in Eq.~\eqref{vor} as
\begin{equation}
\mathbf{U}=\mathbf{U}_0+\mathbf{U}_1+\mathbf{U}_2,
\end{equation}
where the rotational viscous correction $\mathbf{U}_1=(U_1,W_1)$ and the wave-induced irrotational $\mathbf{U}_0$ are  both of $O(a)$. To probe the stability of the laminar state
\begin{equation}
\mathbf{U}_b=\mathbf{U}_0+\mathbf{U}_1,
\end{equation}
we add a disturbance $\mathbf{U}_2=(U_2,W_2)$ of $O(\epsilon)$ smaller than $\mathbf{U}_1$, which is of $O(a)$. Similarly, the vorticity field is expanded as 
\begin{equation}
\omega=\omega_1+\omega_2,
\end{equation}
where $\omega_1$ of $O(a)$ is the vorticity associated with the laminar flow $\mathbf{U}_{b}$ and $\omega_2$ of $O(\epsilon)$ is that of the disturbance $\mathbf{U}_2$. 

The vorticity generated on the zero-stress free surface $z=\eta$ follows from Eq.~\eqref{Omega} as:\begin{equation}
\omega_{s}=-\frac{2\eta_{xx}}{1+\eta_{x}^{2}}\left(u-c\right)=2\eta_{xx}c+O(a^{2}).
\end{equation}
To account for the wavy surface traveling a speed $c$ we make the change of variables $(t=t, X=x-ct, Z=z-\eta)$, and transform the gradient and the Laplacian to
\begin{equation}
\nabla=\left(\partial_{x}+\eta_x\partial_{Z},\partial_{Z}\right),
\end{equation}
and
\begin{equation}
\nabla^{2}=\partial_{xx}+2\eta_{xx}\partial_{xZ}+\left(1+\eta_{xx}^{2}\right)\partial_{ZZ}.
\end{equation}
Then, from Eq.~\eqref{vor} a hierarchy of perturbation equations can be ordered as follows. Neglecting second- and higher-order terms in wave amplitude, the evolution equation of the laminar vorticity is given by
\begin{equation}
O(a)\quad\begin{cases}
\begin{array}{c}
L\omega_{1}=0\\
\\
\omega_{1}(Z=0)=-2\eta_{XX}c,\quad\omega_{1}(Z=-\infty)=0
\end{array}\end{cases},\label{LAM}
\end{equation}
where vorticity is imposed to vanish at infinite depth, and the linear operator
\begin{equation}
L=\partial_{t}-c\partial_{X}-\nu(\partial_{XX}+\partial_{ZZ}).
\end{equation}

The evolution equation of the vorticity disturbance correct to $O(a\epsilon)$ is given by
\begin{equation}
O(a\epsilon)\quad\begin{cases}
\begin{array}{c}
L\omega_2+\mathbf{U}_2\bullet\nabla\omega_1+\left(\mathbf{U}_0+\mathbf{U}_1\right)\bullet\nabla\omega_{2}=0\\
\\
\omega_{2}(Z=0)=0,\quad\omega_{2}(Z=-\infty)=0,
\end{array}\end{cases}\label{OS1}
\end{equation}
where we impose vanishing vorticity at the surface and at infinite depth. Note that the $O(a^2)$ corrections induced by the curvature of the wave surface via the change $Z=z-\eta$ have been neglected. Thus, to $O(a\epsilon)$ the Stokes wave surface is seen as flat. Moreover,  the laminar base flow $U_b$ does not include the Stokes drift 
\begin{equation}
U_{ST}(z)=k a^2\omega\exp(2 k z),\label{Drift}
\end{equation}
since of~$O(a^2)$. 

\subsection{Laminar flow}
Following~\cite{Phillips1977}, we define the Reynolds number $Re=U\delta/\nu$, where $U$ is the characteristic orbital velocity defined above and $\delta$ the boundary layer width, to be determined from the formulation. Then, in the limit of large $Re$, or $\delta\ll a$, we invoke the boundary layer (BL) approximation since within the boundary layer the flow will vary more rapidly with depth, i.e. $\partial_Z\gg\partial_X$. Hence, Eq.~\eqref{LAM} simplifies to
\begin{equation}
O(a)\quad\begin{cases}
\begin{array}{c}
-c\partial_{X}\omega_1-\nu\partial_{ZZ}\omega_1=0\\
\\
\omega_1(Z=0)=-2a\sigma k\cos(kX),\quad\omega_{2}(Z=-\infty)=0
\end{array}\end{cases},
\end{equation} 
where $\partial_t\omega_1$ is set to zero. The laminar correction to the Stokes wave-induced  irrotational flow follows as:
\begin{equation}
\omega_1(X,Z)=-2 U k\exp\left(\frac{Z}{\delta}\right)\cos\left(kX-\frac{Z}{\delta}\right),\label{omega1}
\end{equation}
where the dependence on the time $t$ is intrinsic in the comoving frame $X=x-ct$ moving at the phase speed $c$. Indeed, vorticity is frozen into the flow, which is steady in the comoving frame $X$. The boundary layer width is then explicitly given by
\begin{equation}
\delta=\sqrt{\frac{2\nu}{\sigma}},
\end{equation}
with Reynolds number
\begin{equation}
Re=\frac{U\delta}{\nu}=2\frac{a}{\delta}.\label{Re}
\end{equation}
Note that $Re=\sqrt{2}Re_w^{1/2}$  and the \cite{Babanin2009}'s critical threshold $Re_w\simeq3000$ for transition to turbulence in non-breaking waves corresponds to $Re\approx77$. 

Invoking again the BL approximation, $\omega_1=\partial_X V_1 - \partial_Z U_1\simeq - \partial_Z U_1$ and the associated horizontal velocity of $O(k\delta)$ is given by
\begin{equation}
U_{1}\simeq-\int\omega_{1}dZ=\sqrt{2}U k\delta \exp\left(\frac{Z}{\delta}\right)\cos\left(kX-\frac{Z}{\delta}+\frac{\pi}{4}\right).\label{U1}
\end{equation}
Since $k\delta=2(k a)/Re$, then $U_1$ is of~$O(Re^{-1})$. The associated vertical velocity $W_1$ is neglected since of~$O(Re^{-2})$. However, the irrotational component $W_0$ cannot be neglected.  Note that the flow of a flat boundary layer at the bottom of a progressive or solitary wave is much stronger since its amplitude is of~$O(1)$~\citep{Phillips1977,Blondeaux1979,blondeaux_pralits_vittori2012}.

\subsection{Orr-Sommerfeld equation}
In the limit of large $Re$, or $\delta\ll a$, the laminar base flow in~\eqref{U1} varies rapidly with depth on the scale $\zeta=Z/\delta$ and more gently along the horizontal direction $x$. Drawing on~\cite{Blondeaux1979}, we now assume that the disturbance $\omega_2(\chi,\zeta,\tau)$ also varies vertically as $\zeta$, but it evolves horizontally in space and over time much more rapidly than the base flow, precisely on the fast scales $\chi=x/\delta$ and $\tau=U t/\delta$ respectively. Such an assumption uncouples the fast dynamics of the disturbance from the 'slow' evolution of the unsteady base flow. This allows to examine the 'momentary' stability of the base flow at a given point $x$ in space and at an instant of time $t$, or at the "moment" $X=x-ct$. As a result, we will see that the 'slow' variable $X$ appears as a parameter in the Orr-Sommerfeld equation governing the fast evolution of the disturbance. 

At this stage, we introduce the streamfunction $\Psi(\chi,\zeta,\tau)$ to satisfy the incompressibility of the flow disturbance. On this basis,
\[
U_{2}=\partial_{\zeta}\Psi,\quad W_{2}=-\partial_{\chi}\Psi,
\] 
and  
\[
\omega_2=-\triangle \Psi,
\]
where $\triangle=\partial_{\chi\chi}+\partial_{\zeta\zeta}$. From Eq.~\eqref{OS1}, correct to $O(Re^{-1})$, $\Psi$ satisfies
\begin{equation}
\mathcal{L}\triangle\Psi+\partial_{\chi}\Psi\,\partial_{\zeta\zeta}u_{1}=0,\label{OS2}
\end{equation}
and we assume the simple no-slip boundary conditions
\begin{equation}
\partial_{\chi}\Psi=\partial_{\zeta}\Psi=0,\quad\zeta=0,-\infty
\end{equation}
that impose vanishing velocities at both the surface $\zeta=0$ and at infinite depths. Alternatively, a free-slip $\partial_{\chi}\Psi=\partial_{\zeta\zeta}\Psi=0$ could also be imposed, but we found that the stability of the flow is the same for both kinds of boundary conditions. The operator
\begin{equation}
\mathcal{L}=\partial_{\tau}+\left(u_{0}+u_{1}\right)\partial_{\chi}+w_{0}\,\partial_{\zeta}-\frac{1}{Re}\triangle^2,
\end{equation}
and the dimensionless irrotational velocities
\begin{equation}
u_{0}=U_{0}/U=\exp\left(2\frac{ka}{Re}\zeta\right)\cos(kX),\quad w_{0}=W_{0}/U=\exp\left(2\frac{ka}{Re}\zeta\right)\sin(kX),
\end{equation}
the horizontal viscous component 
\begin{equation}
u_{1}(X,\zeta)=U_{1}/U=U_m\exp\left(\zeta\right)\cos\left(kX-\zeta+\frac{\pi}{4}\right),
\end{equation}
with $U_m=2\sqrt{2}(ka)/Re$ its maximum magnitude. 

Now define the harmonic disturbance
\begin{equation}
\Psi(\chi,\zeta,\tau;X)=G(\zeta;X)\exp\left[i\alpha(\chi-C(X)\tau)\right]
\end{equation}
with streamwise wavenumber $\alpha$ and complex velocity $C(X)=C_r(X)+iC_i(X)$ in a moving frame with the Stokes wave. Note that $C(X)$ depends on the slow scale of the base flow. Here, $X=x-ct$ is just a parameter that selects the 'moment' at which stability is probed.  The real part $C_r$ is the phase speed of the disturbance and the imaginary part $C_i$ is the associated growth rate. If $C_i>0$ the disturbance will grow in time leading to instability. At a 'moment' $X$, the eigenmode $G(\zeta;X)$ associated with $C(X)$ yields the vertical structure of the disturbance. From Eq.~\eqref{OS2} it satisfies the Orr-Sommerfeld (O-S) eigenvalue problem  
\begin{equation}
\mathsf{L}G=0,\label{OS}
\end{equation}
defined on the semi-infinite domain $\zeta\in\left[0,-\infty\right)$ with no-slip boundary conditions
\begin{equation}
\Psi=\partial_{\zeta}\Psi=0, \quad\zeta=0,-\infty,\label{BC}
\end{equation}
and the OS operator
\begin{equation}
\mathsf{L}=\left(-C(X)+u_{0}+u_{1}\right)\mathsf{N\textrm{+\ensuremath{\partial_{\zeta\zeta}u_{1}}}}+w_{0}\,\partial_{\zeta}\mathsf{N}-\frac{1}{i\alpha Re}\mathsf{N}^{2},\label{OSop}
\end{equation}
and $\mathsf{N}=-\alpha^{2}+\partial_{\zeta\zeta}$. Here, $u_0$,$u_1$ and $w_0$ depend on the parameter $X$.

\begin{figure}[h]
\centering\includegraphics[scale=0.64]{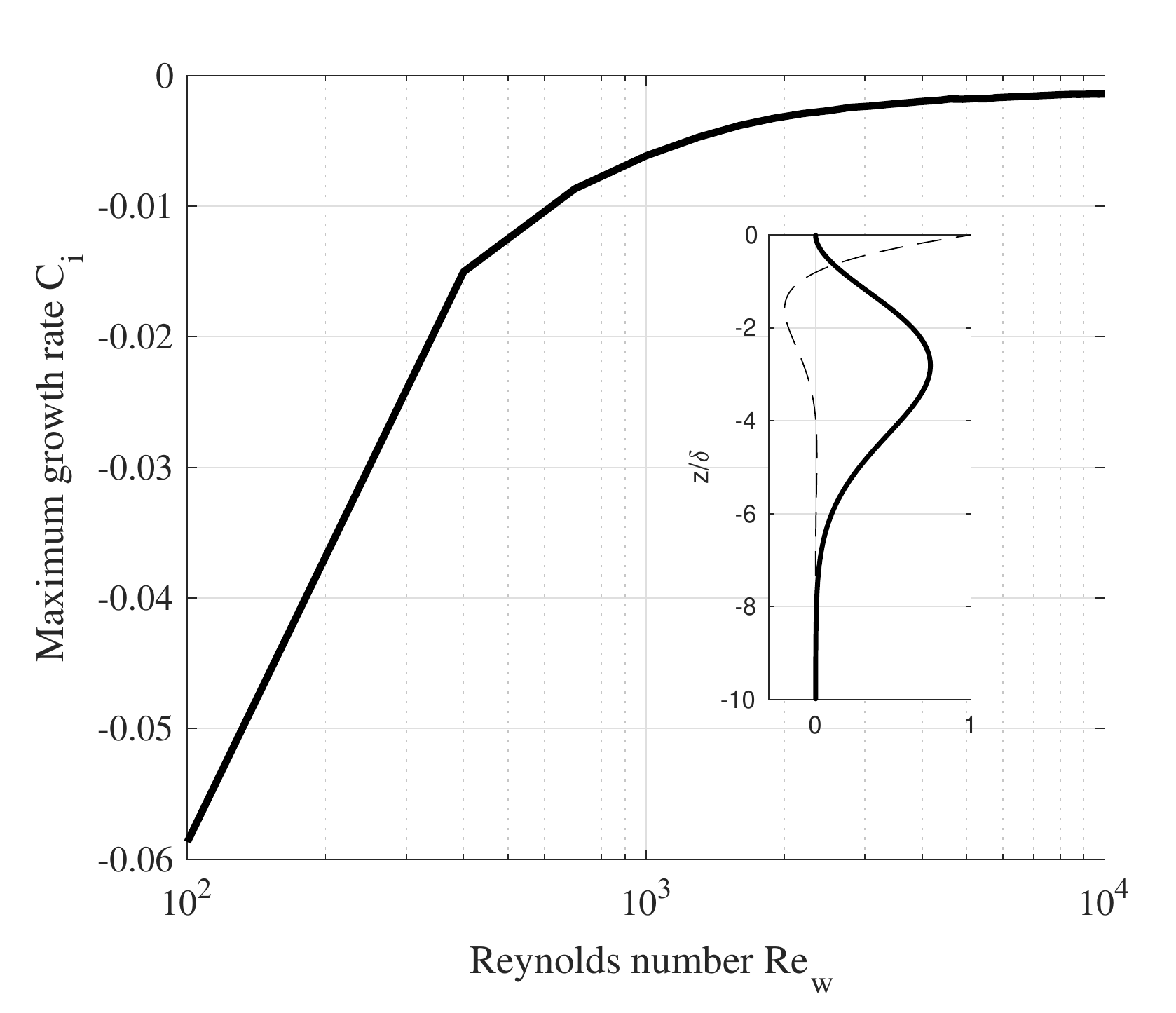} 
\caption{Maximum growth rate $C_i$ of a 2-D perturbation as a function of the wave Reynolds number $Re_w$ for wavenumber $\alpha=1.5$ and wave steepness $ka=0.35$. The inset depicts the typical stable eigenmode and the viscous component of the base Stokes flow. }
\label{FIG1}
\end{figure}

\begin{figure}[h]
\centering\includegraphics[scale=0.64]{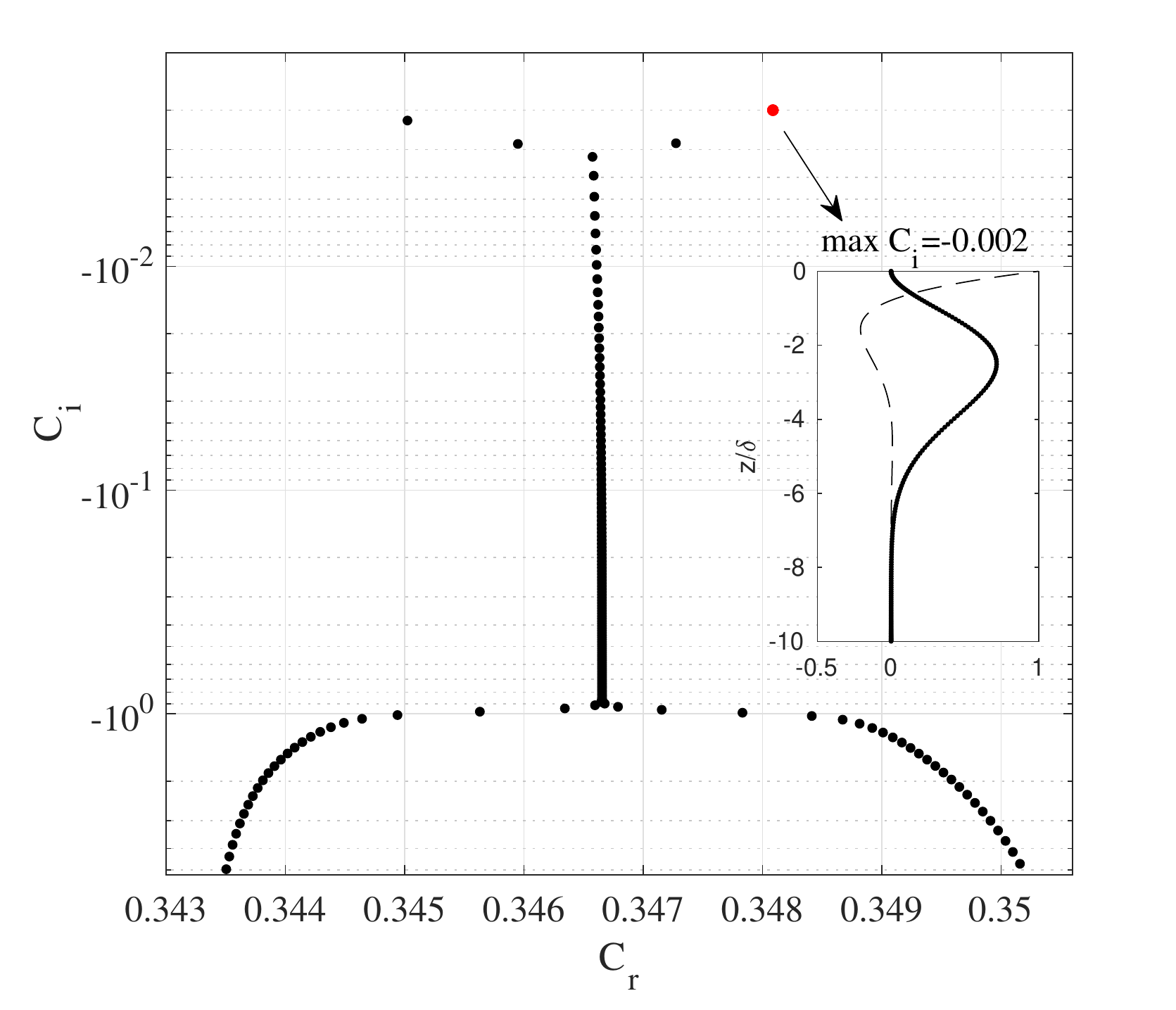} 
\caption{O-S Spectrum of a 2-D perturbation for $Re_w=3000$ at the moment $X=0$ of maximum growth, wavenumber $\alpha=1.5$ and wave steepness $ka=0.35$. The inset depicts the least stable eigenmode and the viscous component of the base Stokes flow.}
\label{FIG2}
\end{figure}

\subsection{Orr-Sommerfeld Spectrum}

Drawing on~\cite{trefethen2000spectral}, we approximate the O-S operator in~\eqref{OSop} using a spectral Chebyshev discretization. The stability problem in~\eqref{OS} is defined on the semi-infinite domain $\zeta\in\left[0,-\infty\right)$. The computational domain is then set in $[-\zeta_{max}, 0]$, where the depth $\zeta_{max}$ is referred to as the 'depth of influence' since it marks the perceptible depth of penetration of vorticity~\citep{leibovich_radhakrishnan_1977}. It is determined by ensuring that in the region outside the BL vorticity and its gradient smoothly decay to zero imposed at  $\zeta_{max}$. For the parameter range investigated we found the optimal value $\zeta_{max}=10.5$, or equivalently $10.5\delta$, with $\delta$ the boundary layer width. Preceding discretization, the computational domain is mapped onto the Chebyshev interval $\hat{\zeta}\in[-1,1]$ . To do so, we use the rational map~\citep{Henningson2001} 
\[
\zeta=f(\hat{\zeta})=a\frac{1+\hat{\zeta}}{b-\hat{\zeta}},\qquad a=\frac{\zeta_{i}\zeta_{max}}{\zeta_{max}-2\zeta_{i}},\;b=1+\frac{2a}{\zeta_{max}}
\]
which clusters grid points more densely near the boundary layer region to have it sufficiently resolved. In particular, this mapping distributes half of the grid points in the interval $0>\zeta>\zeta_i$ and the other half in the rest of the computational domain. Here, we use $\zeta_i=5$, or equivalently $5\delta$. 
Then, the streamfunction $\Psi$ is approximated as
\[
\Psi(\hat{\zeta})\simeq\sum_{k=0}^{N}a_{k}T_{k}(\hat{\zeta})(1-\hat{\zeta}^{2}),\qquad\hat{\zeta}\in\left[-1,1\right],
\]
where $T_{k}(\hat{\zeta})=\cos(k\,\arccos\hat{\zeta})$ is the $k^{th}$ Chebyshev polynomial of the first kind, $a_{k}$ the  corresponding amplitude, and the factor $(1-\hat{\zeta}^{2})$ forces the no-slip boundary conditions~\eqref{BC}~\citep{trefethen2000spectral}. For slip boundary conditions that factor is replaced by $(1-\hat{\zeta}^{2})^2$. 
Chebyshev discretization converts $\mathsf{L}G=0$ in~\eqref{OS} into the generalized eigenvalue problem
\[
 \mathbf{A}\mathbf{w}=C(X)\mathbf{B}\mathbf{w},
 \]
 where  $\mathbf{A}$ and $\mathbf{B}$ are matrices of size $N\times N$, $\mathbf{w}$ is the $(N\times 1)$-eigenvector listing the discretized values of the eigenfunction $G(\zeta=f(\hat{\zeta}))$ at the Gauss-Lobatto points of the Chebyshev domain~\citep{trefethen2000spectral}, and $C(X)$ is the associated eigenvalue. 

We target the least stable/unstable eigenvalues, which are closest to the origin of the complex plane. These are obtained with a 2-digit accuracy for a moderate value of $N=200$, since small eigenvalues tend to correspond to smoother eigenmodes, resolvable on coarser grids~\citep{Trefethen2005}. So, the eigenvalues can be computed directly without resorting to Krylov methods~\citep{Trefethen2005}. 

The maximum growth rate $C_i$ observed over the entire range of moments $X$ in $[-\pi,\pi]$ is plotted in Fig. 1 as a function of the wave Reynolds number $Re_w$. The 2-D perturbation has a wavenumber $\alpha=1.5$ and the wave steepness $ka=0.35$. No instability is observed at the proposed critical $Re_w=3000$ as the base flow is linearly stable in the investigated range of Reynolds numbers up to $Re_w=10^4$. The inset in the same figure depicts the typical shape of the least stable eigenmode and the viscous component of the Stokes flow. Note that the eigenmode attains its maximum outside the boundary layer and its vorticity is practically null at the two ends of the domain.

The O-S Spectrum is shown in Fig. 2  at the moment when the crest of the Stokes wave travels across $x=0$ at time $t=0$, i.e. the moment $X=0$ of maximum growth. We observe a Y-shaped distribution and spurious modes for $C_i<-10$, whose eigenfunctions are localized at the  bottom $\zeta=-\zeta_{max}$ of the computational domain. Since the fluid motion takes place in an infinite domain, the spectrum includes a discrete as well as a continuous part, whose discrete representation is given by the vertical branch of the Y-shaped form. The inset depicts the shape of the least stable eigenmode and the viscous component of the base Stokes flow. The eigenmode peaks outside the laminar boundary layer. 





\subsection{Orr-Sommerfeld pseudospectrum}

The O-S operator in~\eqref{OSop} is non-normal since $\mathsf{L}$ is not equal to its adjoint $\mathsf{L}^*$, and its eigenfunctions are not orthogonal~\citep{Trefethen2005}. The associated Chebyshev matrix $\mathbf{F}=\mathbf{B}^{-1}\mathbf{A}$ inherits the same  non-normality as $\mathbf{F}\neq\mathbf{F}^*$, where $\mathbf{F}^*$ is the conjugate transpose of $\mathbf{F}$.

The behavior of a normal matrix or operator is completely determined by its eigenvalues. On the contrary, non-normal matrices and operators have their eigenvalues highly sensitive to perturbations, and the eigenfunctions, though complete, may be nearly linearly dependent~\citep{Trefethen2005}. Furthermore, even if the base laminar flow is stable, a small perturbation added to it undergoes an energy transient growth before monotonically decaying to vanishing amplitudes over large times~\citep{Trefethen1993,Henningson2001,Fedele_pipe2005}. As a result, the base flow is prone to become unstable to finite amplitude perturbations~\citep{Henningson2001}.

Assume that we alter the laminar base flow by adding finite imperfections, or defects, and wish to investigate the stability of the imperfect base flow. For example, we can artificially introduce small to finite defects into the base laminar flow by artificially altering (increasing) the magnitude $U_m$ of the viscous component $u_1$ up to $O(1)$~finite imperfections. We observe that the stable eigenvalues of the O-S spectrum of the original (perfect) base flow can change by a finite amount and become unstable because of the added imperfections.  This is seen in Fig. 3  which shows the maximum growth rate $C_i$ observed over the entire range of moments $X$ in $[-\pi,\pi]$ as a function of the amplitude $U_m$ for wavenumber $\alpha=1.5$ and wave steepness $ka=0.35$. The imperfect base flow becomes unstable for $U_m>0.2$ and the associated eigenmodes attain their maximum within the boundary layer, whereas the stable eigenmodes peak beneath it. Note that when $U_m$ is of~$O(1)$ the laminar flow is similar to that at the bottom of a progressive or solitary wave in finite depth, which is known to be unstable~\citep{Blondeaux1979,blondeaux_pralits_vittori2012}. 
Another way to make the base flow imperfect is by adding a finite amplitude Stokes drift $U_{ST}$ (see Eq.~\eqref{Drift}). Nevertheless, our analysis indicates that it does not affect the stability of the flow. Indeed, as discussed later, the flow may likely be unstable to streamwise vortical rolls similar to the Langmuir cells generated by the Stokes and wind drifts~\citep{craik_leibovich1976,leibovich_1977,leibovich_radhakrishnan_1977,leibovich1980,leibovich_paolucci_1981}. 

The sensitivity of the O-S operator to imperfections, or defects of the base flow can be rigorously quantified by means of the pseudospectrum of the matrix $\mathbf{F}$~\citep{Trefethen2005,Trefethen1999,Henningson2001}. 
Define the 2-norm of a matrix $\mathbf{E}$ as $\left\Vert \mathbf{E}\right\Vert =\sigma_{\mathrm{max}}(\mathbf{E})$, where $\sigma_{\mathrm{max}}(\mathbf{E})$ is the maximum singular value of $\mathbf{E}$, i.e. the maximum eigenvalue of $\mathbf{E}^*\mathbf{E}$. Then, the pseudospectrum is defined as
\begin{equation}
\varLambda_{\epsilon}(\mathbf{F})=\left\{ z\in\mathbb{C:\mathit{z\in\varLambda(\mathbf{F+E})\:\,\mathrm{for\,\,some}\,\,}}\mathbf{E}\,\,\mathrm{with}\,\,\left\Vert \mathbf{E}\right\Vert \leq\epsilon\right\}, 
\end{equation}
the set of all the complex numbers that are in the spectrum of some matrix $\mathbf{F+E}$ obtained by altering  $\mathbf{F}$ by a perturbation $\mathbf{E}$ of norm at most $\epsilon$.  Thus, pseudospectra can be interpreted in terms of perturbations of spectra. Note that $\varLambda_{0}(\mathbf{F})=\varLambda(\mathbf{F})$ is the sets of eigenvalues of $\mathbf{F}$. 

For computational purposes the pseudospectrum  can be defined equivalently as~\citep{Trefethen2005}
\begin{equation}
\varLambda_{\epsilon}(\mathbf{F})=\left\{ z\in\mathbb{C:\sigma_{\mathrm{min}}\left(\mathbf{\mathit{z}\mathbf{I}-F}\right)}\leq\epsilon\right\}, 
\end{equation}
where $\sigma_{\mathrm{min}}$ denotes the minimum singular value of $\mathbf{\mathit{z}\mathbf{I}-F}$ and the associated singular vector  is the pseudo-eigenmode at $z$. Thus, the pseudospectrum of $\mathbf{F}$ is made up of all the sets in the complex plane bounded by the $\epsilon-$~level curves of the function $\sigma_{\mathrm{min}}(z\mathbf{I}-\mathbf{F})$. 

If $\mathbf{F}$ were normal, $\varLambda_{\epsilon}(\mathbf{F})$ would be the union of the $\mathbf{E}$-balls centered at the eigenvalues. Actually, $\mathbf{F}$ is nonnormal and its pseudospectrum is much larger than the O-S spectrum. This is clearly seen in the central panel of Fig 4, which depicts the pseudospectrum $\varLambda_{\epsilon}(\mathbf{F})$ at the moment $X=x-ct=0$ for $Re_w=3000$, $\alpha=1.5$ and $ka=0.35$. The dots represent all the stable eigenvalues of the associated O-S spectrum. The contour lines denote where the eigenvalues would lie if the matrices approximating the O-S operator were perturbed by an amount $\epsilon$. In  particular, matrix perturbations of order $\epsilon\~O(10^{-1})$ lead to significant changes in the unperturbed stable eigenvalues that become unstable. Such small changes in the O-S matrices correspond to altering the amplitude $U_m$ of the base flow to $O(1)$, which becomes unstable in agreement with the instability shown  in Fig. 3. An unstable pseudo-eigenmode  is shown in the right panel of Fig. 4.  Note that it attains its maximum within the laminar boundary layer. 



In summary, the laminar boundary layer flow beneath a Stokes wave is linearly stable. However, it is structurally unstable since finite imperfections added to the base flow can trigger instability similar to what has been observed in channel flows~\citep{Waleffe1995,WaleffePOF1997,FaisstEckhardt}. Clearly, the vortex stretching or tilting of the perturbation by the base flow is absent since the model is 2-D. To account for such effects, in the next section we will investigate the stability of the laminar flow to 3-D streamwise-independent perturbations. 

\begin{figure}[h]
\centering\includegraphics[scale=0.64]{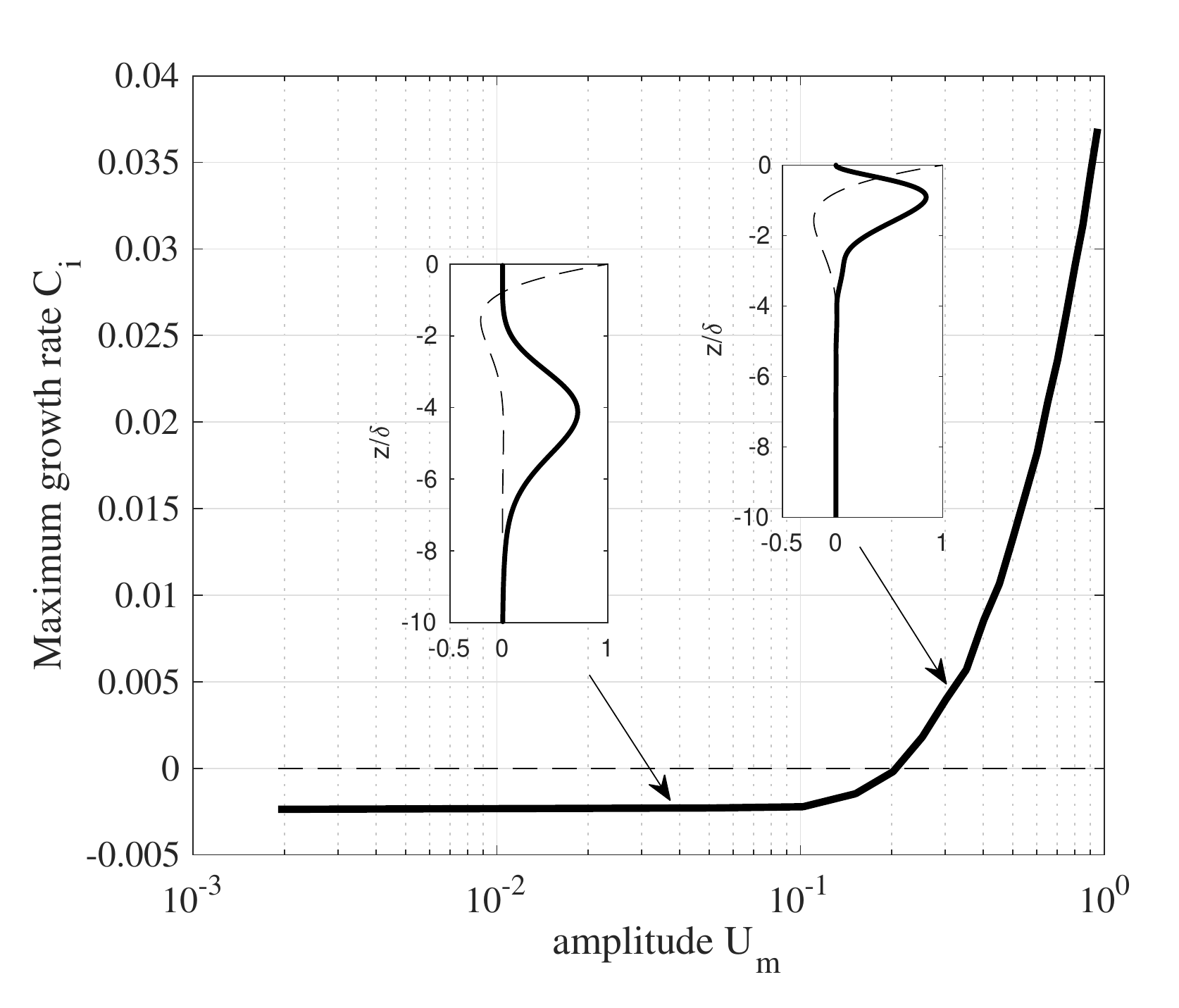} 
\caption{Maximum growth rate $C_i$ of a 2-D perturbation to an imperfect base flow as a function of the amplitude $U_m$ for $Re_w=3000$, wavenumber $\alpha=1.5$ and wave steepness $ka=0.35$. The typical shapes of stable and unstable eigenmodes and the viscous component of the base flow are also shown. }
\label{FIG3}
\end{figure}

\begin{figure}[h]
\centering\includegraphics[scale=0.64]{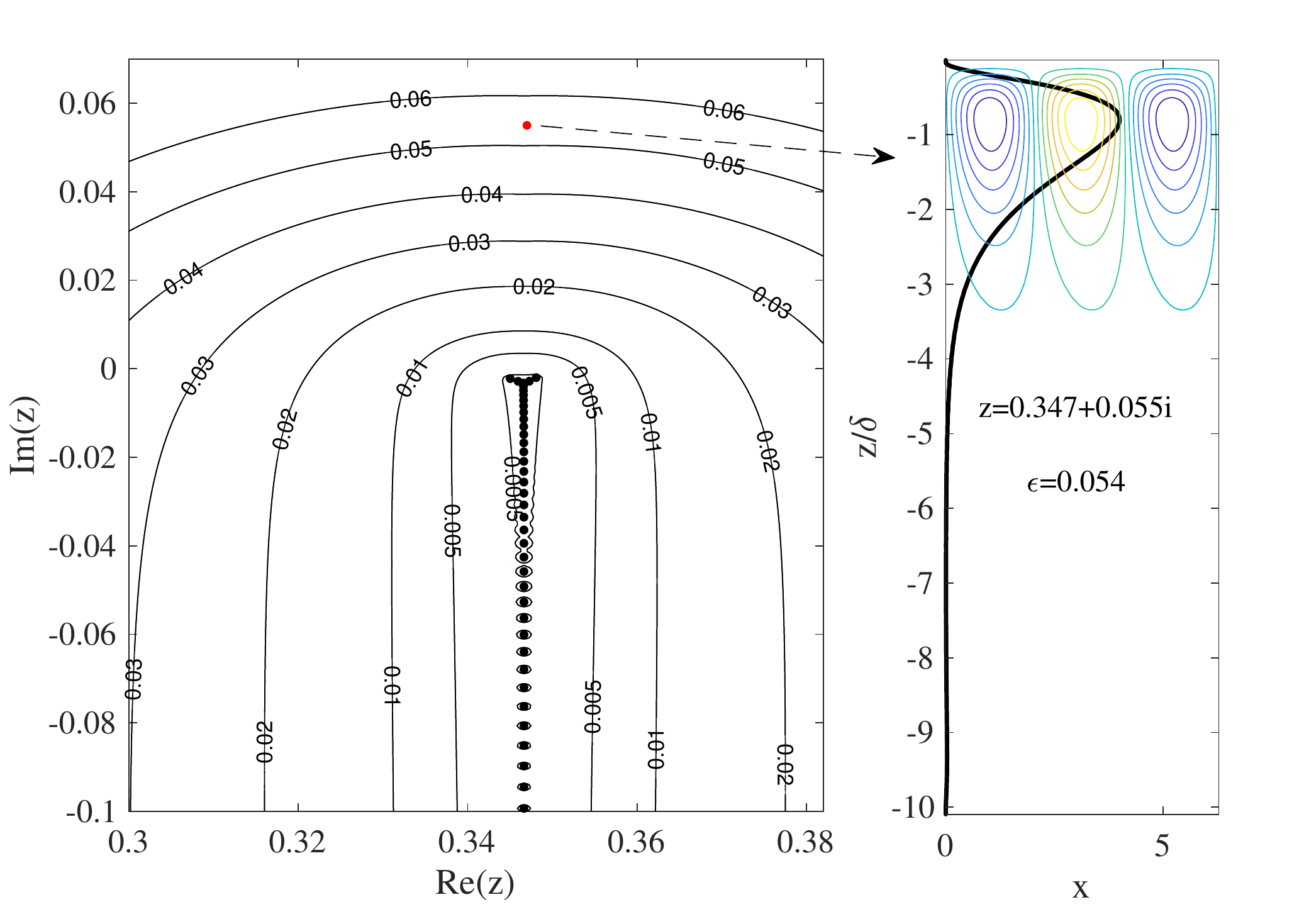} 
\caption{(central panel) Pseudospectrum for a 2-D perturbation at the moment $X=x-ct=0$ for $Re_w=3000$, $\alpha=1.5$ and $ka=0.35$. (Right panel) An unstable pseudoeigenmode at $z=0.347+0.055i$ (red dot) and $\epsilon=0.054$.}
\label{FIG4}
\end{figure}

\section{Stability to 3-D streamwise-independent perturbations}

Consider the $O(a)$~base Stokes flow and associated vorticity
\begin{equation}
\mathbf{U}_{b}(X,\zeta)=(U_{b},0,W_{b}),\qquad\mathbf{\boldsymbol{\omega}}_{b}(X,\zeta)=(0,\omega_{b}=-\partial_{\zeta}U_{1},0),
\end{equation}
where $U_{b}=U_0+U_1$, $W_b=W_0+W_1$, and $\omega_b=\omega_1$ from Eq.~\eqref{omega1}.
Perturb the base flow by a generic 3D perturbation of~$O(\epsilon)$ 
\begin{equation}
\boldsymbol{U}_{2}(x,y,z,t)=(U_{2},V_{2},W_{2})
\end{equation}
with vorticity $\boldsymbol{\omega}_{2}(x,z,t)=(\omega_{x},\omega_{y},\omega_{z})$. The dynamics of the perturbation is governed by the coupled equations
\begin{equation}
\left\{ \begin{array}{c}
\partial_{t}\omega_{x}+\left((\mathbf{U}_{b}+\mathbf{U}_{2})\bullet\nabla\right)\omega_{x} =\left((\mathbf{\boldsymbol{\omega}}_{b}+\boldsymbol{\omega}_{2})\bullet\nabla\right)(U_{b}+U_{2})+\nu\nabla^{2}\omega_{x}\\
\\
\partial_{t}U_{2}+\left((\mathbf{U}_{b}+\mathbf{U}_{2})\bullet\nabla\right)(U_{b}+U_{2})=-\frac{1}{\rho}\partial_{x}p+\nu\nabla^{2}(U_{b}+U_{2}),
\end{array}\right.\label{3DNS} 
\end{equation}
for the streamwise vorticity $\omega_x$ and velocity $U_2$, and $\nabla\bullet\mathbf{U}_{2}=0$ and $p$ is pressure. 
We now assume that the flow disturbance and pressure are streamwise independent and drop the dependence on $x$ in Eqs.~\eqref{3DNS}. 

Similar to the 2-D stability, we draw on~\cite{Blondeaux1979} and assume that the disturbance $\boldsymbol{U}_{2}(\xi,\zeta,\tau)$ varies vertically as fast as the BL scale, i.e. $\zeta=z/\delta$, but that it evolves spanwise in space and over time much more rapidly than the base flow, i.e. on the fast scales $\xi=y/\delta$ and $\tau=U t/\delta$ respectively. The fast dynamics of the disturbance is then uncoupled from the slow evolution of the unsteady base flow. The 'momentary' stability of the base flow at a given point $x$ in space and at an instant of time $t$, or at the "moment" $X=x-ct$ can be probed, and the slow variable $X$ appears as a parameter in the stability equations shown below. Then, the vorticity of the perturbation $\boldsymbol{U}_{2}$ follows as
\begin{equation}
\boldsymbol{\omega}_{2}(\xi,\zeta,\tau)=(\omega_{x},\omega_{y},\omega_{z})=(\partial_{\xi}W_{2}-\partial_{\zeta}V_{2},\partial_{\zeta}U_{2},-\partial_{\xi}U_{2}),
\end{equation}
and
\begin{equation}
\left(\boldsymbol{\omega}_{2}\bullet\nabla\right)U_{2}=0,
\end{equation}
implying that vortex stretching is absent. However, the flow perturbation is stretched and tilted by the base flow as $\left(\boldsymbol{\omega}_{b}\bullet\nabla\right)U_{2}$ does not vanish.
The required flow incompressibility constraint 
\begin{equation}
\partial_{\xi}V_{2}+\partial_{\zeta}W_{2}=0\label{div}
\end{equation}
is satisfied a priori by defining a streamfunction $\Psi(\eta,\xi,\tau)$. Then, the spanwise velocity components follow as
\begin{equation}
V_{2}=\partial_{\zeta}\Psi,W_{2}=-\partial_{\xi}\Psi,
\end{equation}
and vorticity $\omega_{x}=\partial_{\xi}W_{2}-\partial_{\zeta}V_{2}=-\triangle\Psi$. 

Correct to $O(Re^{-1})$, or $O(\delta)$, the linearized stability equations for the 3-D streamwise-independent disturbance follow from Eqs.~\eqref{3DNS} as
\begin{equation}
\left\{ \begin{array}{c}
\left(\partial_{\tau}+w_0\partial_{\zeta}-\frac{1}{Re}\triangle-2 (k\delta)\partial_{\widetilde{X}}u_0\right)\triangle\Psi-\partial_{\zeta}(2 u_1+u_0)\partial_{\xi}U_{2}=0\\
\\
\left(\partial_{\tau}+w_0\partial_{\zeta}-\frac{1}{Re}\triangle+2 (k\delta)\partial_{\widetilde{X}}u_0\right)U_{2}-\partial_{\zeta}(u_0+u_1)\partial_{\xi}\Psi=0
\end{array}\right.,\label{3DOS}
\end{equation}
with no-slip boundary conditions
\begin{equation}
\partial_{\zeta}\Psi=\partial_{\xi}\Psi=U_{2}=0,\qquad\zeta=0,-\infty,\label{bc}
\end{equation}
where $\widetilde{X}=kX$, $\triangle=\partial_{\xi\xi}+\partial_{\zeta\zeta}$ and $u_0=U_0/U,w_0=W_0/U$. Note that in the current linearized model, correct to $O(a\epsilon)$, the laminar base flow does not include the Stokes drift of $O(a^2)$. Interestingly, if we neglect the $X$-dependence of the base flow and set the Langmuir number $La=Re^{-1}$ as the inverse of the Reynolds number, then Eqs.~\eqref{3DOS} coincide with the linearized version of the Langmuir model of~\cite{leibovich_1977}. In his model the flow beneath the slow wind drift evolves on fast scales and has also streamwise-independent velocities. The \cite{leibovich_1977}~model predicted the subsurface vortex motions induced by Langmuir cells or rolls, which generate streaks visible at the surface of the ocean~\citep{craik_leibovich1976}.

Consider now the ansatz
\begin{equation}
\Psi=G(\zeta)\exp\left[i\beta(\xi-C(X)\tau)\right],\quad U_{2}=F(\zeta)\exp\left[i\beta(\xi-C(X)\tau)\right],
\end{equation}
with streamwise wavenumber $\beta$ and complex velocity $C(X)=C_r(X)+iC_i(X)$. Here, $X=x-ct$ selects the 'moment' at which stability is probed.  $C_r$ is the phase speed of the disturbance and $C_i$ is the  associated growth rate.  At a 'moment' $X$, the eigenmodes $G(\zeta;X)$ and $F(\zeta;X)$ associated with $C(X)$ yield the vertical structure of the disturbance, and from Eq.\eqref{3DOS} they satisfy the eigenvalue problem
\begin{equation}
\left[\begin{array}{cc}
\left( \mathsf{L}-(k\delta)\partial_{\tilde{X}}u_0\right)\mathsf{N} & -i\beta\,\partial_{\zeta}(2u_1+u_0)\\
\\
i\beta\,\partial_{\zeta}(u_0+u_1) &  \mathsf{L}+(k\delta)\partial_{\tilde{X}}u_0
\end{array}\right]\left[\begin{array}{c}
G\\
\\
F
\end{array}\right]=\mathbf{0}\label{LS}
\end{equation}
and boundary conditions
\begin{equation}
G=F=\partial_{\zeta}G=\partial_{\zeta}F=0,\quad\zeta=0,-\infty,\label{bc1}
\end{equation}
where the differential operators
\begin{equation}
\mathbf{\mathsf{\mathsf{L}}}=-i\beta\,C(X)+w_0\partial_{\zeta}-\frac{1}{R_{e}}\mathsf{N}
\end{equation}
and
\begin{equation}
\mathsf{N}=-\beta^{2}+\partial_{\zeta\zeta}.
\end{equation}


Our analysis reveals that the base Stokes flow is also stable to such 3-D perturbations, but it is structurally unstable. Indeed, if we alter the base flow by adding a defect or imperfection, the flow becomes unstable. Without finite defects the base flow is stable. In particular, the pseudospectrum at the moment $X=x-ct=0$ is depicted in the left panel of Figure 5 for $Re_w=3000$, $\beta=1.5$ and $ka=0.35$. The dots represent the stable eigenvalues of the O-S spectrum for the base Stokes flow. The spanwise streamfunction $\Psi$ of an unstable pseudo-eigenmode is shown in the right panel of Figure 5. It is made of streamwise rolls similar to Langmuir cells~\citep{leibovich_1977}.

\begin{figure}[h]
\centering\includegraphics[scale=0.6]{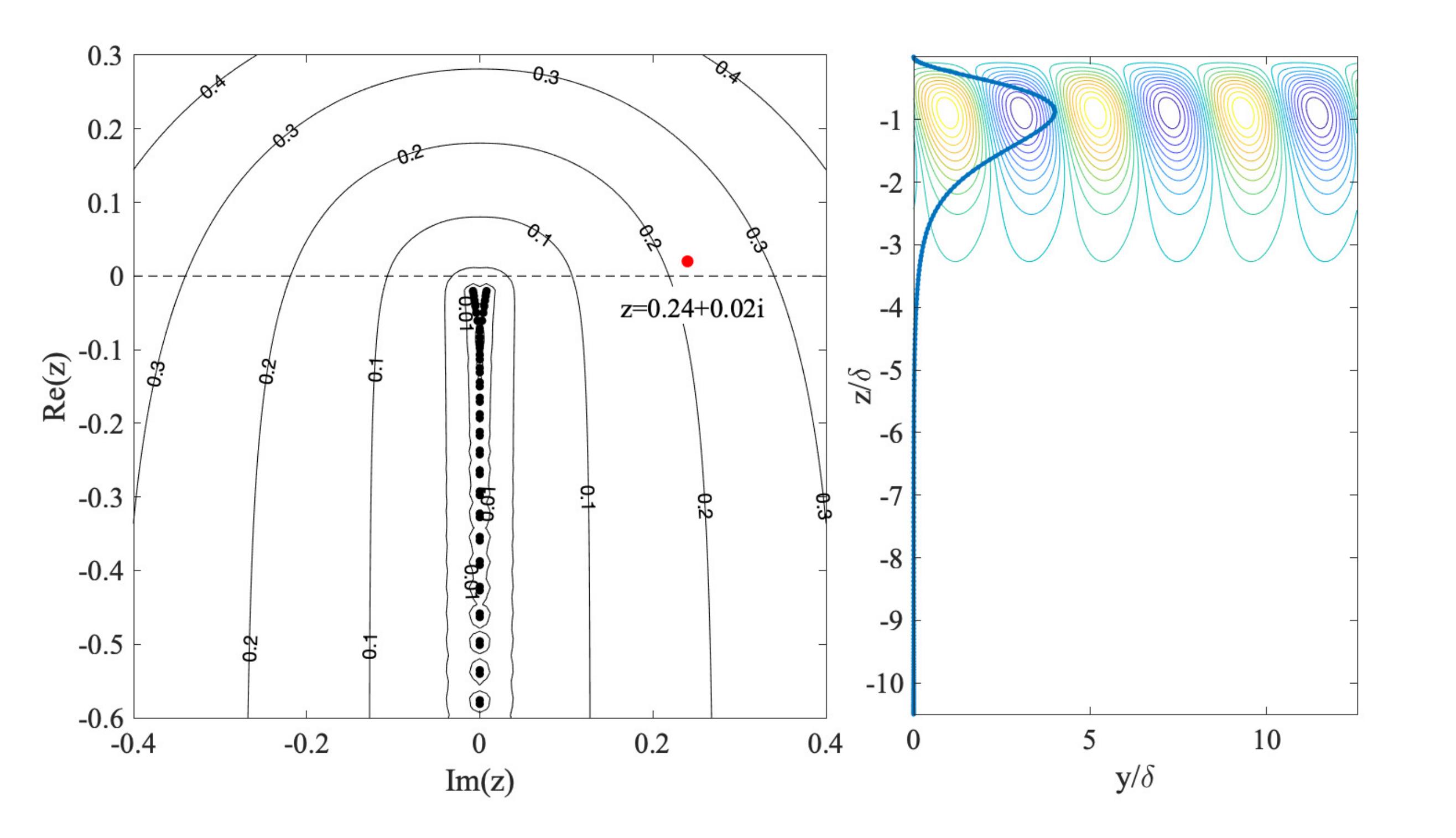} 
\caption{(Left panel) Pseudospectrum for a 3-D perturbation at the moment $X=x-ct=0$ for $Re_w=3000$, $\beta=1.5$ and $ka=0.35$.  (Right panel)  Unstable pseudoeigenmode at $z=0.24+0.02i$ (red dot) and $\epsilon=0.22$.}
\label{FIG5}
\end{figure}

\section{Conclusions}

Our analysis indicates that the unsteady flow of the laminar boundary layer beneath a Stokes wave is linearly stable to 2-D and 3-D streamwise-independent infinitesimal perturbations up to $Re_w\simeq10^4$ for no-slip or free-slip boundary conditions. Thus, it is very likely that the laminar flow is linearly stable at any Reynolds number, similar to Poiseuille flows~\citep{Trefethen1999,Henningson2001}. However, the laminar flow is structurally unstable since it is sensitive to finite imperfections added to it, because of the non-normality of the O-S equations. Moreover, the model equations for 3-D streamwise-independent perturbations coincide with those of the~\cite{leibovich_1977} Langmuir cell model. In particular, an analysis of the associated pseudospectrum reveals that a base flow with imperfections becomes unstable, and unstable pseudo-eigenmodes  shape like streamwise vortical rolls similar to Langmuir cells~\citep{leibovich_1977}. 

Thus, it is plausible that the laminar boundary layer of a Stokes wave is nonlinearly unstable to finite perturbations, similar to the wall boundary layer of channel flows~\citep{Trefethen1999,Henningson2001}. Similar to the role of Waleffe's SSPs in capturing the essence of turbulence in channel flows, Langmuir-type cells could represent the SSP structures dominant in the transition to turbulence by unforced non-breaking waves. To date the author is not aware of any measurements reporting the highlighted circulation or the ambient flow during the turbulent bursts and mixing in the subsurface of non-breaking waves. 

Moreover, the experimental observation of SSPs in pipe flows support the relevance of these unstable states in capturing the nature of fluid turbulence~\citep{HofScience2004}. This also suggests that the chaotic dynamics of Navier-Stokes flows can be effectively unveiled by exploring the state space of associated  high-dimensional dynamical  system~\citep{GibsonetalJFMCouette2008,WillisCvitanovic2013,budanur2017}. Here, turbulence is viewed as an effective random walk in state space through a repertoire of invariant solutions of the governing equations~\citep{CvitanovicJFM2013_clockwork,ChaosBook}. In state space, turbulent trajectories visit the neighbourhoods of equilibria, travelling waves or periodic orbits, switching from one saddle to another through their stable and unstable manifolds~\citep{CvitanovicPOT_1991}. These studies present evidence that unstable periodic orbits provide the skeleton underpinning the chaotic dynamics. 

Thus, the similarity of Langmuir and SSPs structures suggests that a dynamical systems approach to the nonlinear instability of the free surface boundary layer flow in a deep-water Stokes wave will shed more light into the nature of upper ocean turbulence.

\section{Acknowledgements}
FF thanks Profs. Michael Banner and Aziz Tayfun for useful comments and discussions. 

\section{Declaration of Interests}
The author reports no conflict of interest.

\bibliographystyle{jfm}
\bibliography{biblio-full}

\end{document}